\documentclass[authorversion, nonacm]{acmart}
\AtBeginDocument{%
  \providecommand\BibTeX{{%
    \normalfont B\kern-0.5em{\scshape i\kern-0.25em b}\kern-0.8em\TeX}}}
\usepackage{fancyhdr}
\begin{document}
  
\setcopyright{none}

\title{Regulating AI-Based Remote Biometric Identification. Investigating the Public Demand for Bans, Audits, and Public Database Registrations}

\author{Kimon Kieslich}
%\authornote{}
\email{k.kieslich@uva.nl}
\orcid{0000-0002-6305-2997}
\affiliation{%
  \institution{University of Amsterdam}
  \city{Amsterdam}
  \country{Netherlands}
}

\author{Marco Lünich}
\email{marco.luenich@hhu.de}
\orcid{0000-0002-0553-7291}
\affiliation{%
  \institution{Heinrich Heine University Düsseldorf}
  \city{Düsseldorf}
  \country{Germany}}

\renewcommand{\shortauthors}{Kieslich \& Lünich}
\renewcommand{\shorttitle}{Regulating AI-Based Remote Biometric Identification}
\begin{abstract}
AI is increasingly being used in the public sector, including public security. In this context, the use of AI-powered remote biometric identification (RBI) systems is a much-discussed technology. RBI systems are used to identify criminal activity in public spaces, but at the same time they are criticised for inheriting biases and violating fundamental human rights. As a result, the use of RBI poses risks to society. It is therefore important to ensure that such systems are developed in the public interest, which means that any technology that is deployed for public use needs to be scrutinised. While there is a broad consensus among business leaders, policymakers and scientists that AI must be developed in an ethical and trustworthy manner, scholars have argued that ethical guidelines do not guarantee ethical AI, but rather prevent stronger regulation of AI for the Common Good. As a possible counterweight, public opinion can have a decisive influence on policymakers (e.g. through voter demands) to establish boundaries and conditions under which AI systems should be used -- if at all. However, we know little about the conditions that lead to regulatory demand for AI systems. In this study, we focus on the role of trust in AI as well as trust in law enforcement as potential factors that may lead to demands for regulation of AI technology. In addition, we explore the mediating effects of discrimination perceptions regarding RBI. We test the effects on four different use cases of RBI varying the temporal aspect (real-time vs. post hoc analysis) and purpose of use (persecution of criminals vs. safeguarding public events) in a survey among German citizens. We found that German citizens do not differentiate between the different modes of application in terms of their demand for RBI regulation. Furthermore, we show that perceptions of discrimination lead to a demand for stronger regulation, while trust in AI and trust in law enforcement lead to opposite effects in terms of demand for a ban on RBI systems. 
\end{abstract}




\pagestyle{fancy}
\fancyfoot[R]{%
  \ifnum\value{page}>1 % Check if page number is greater than 1
    To be published at ACM FAccT'24 % Your comment goes here
  \fi
}
\maketitle

\hypertarget{introduction}{%
\section{Introduction}\label{introduction}}

The European AI Act marks a key regulatory milestone in the regulation
of AI. The AI Act guarantees that AI technologies must be classified
into different risk classes \cite{EuropeanCommission.2021}. It also includes a class of
non-acceptable risk technologies. Systems in this class have too
far-reaching negative risks for society and/or conflict with the EU's
core ethical values. For example, social scoring systems should be
banned, as well as, with some exceptions, \emph{real-time} biometric
remote identification (RBI) systems \cite{EuropeanParliament.2023}. RBI systems analyse
biometric data (e.g., faces, fingerprints) to identify individuals. The
data is usually collected from video footage, such as surveillance
cameras in public areas. Law enforcement agencies, in particular, are
eager to use RBI to identify criminals or find missing persons \cite{vanBrakel.2011}.
However, the use of RBI is criticized for standing in conflict with
fundamental rights, for instance, in being an unnecessary invasion of
people's privacy and for discrimination of citizens \cite{Barkane.2022, Neuwirth.2023, Vogiatzoglou.2019}.

In the European Parliament, the classification of real-time RBI as an
unacceptable risk technology has led to some discussions with
conservative parties arguing against a ban, as RBI would be useful for
strengthening domestic security \cite{EPPGroup.2023}. The regulation of RBI systems
was also a heated discussion point in the final negotiations about the
end-version of the EU AI Act. While these discussions take place in
political institutions, the opinion of citizens on RBI are also
relevant, as the European Parliament and European Council as democratic
representative institutions need to heed the demands of citizens to a
certain extent.

With the ever-growing influence of AI on society, how citizens want to
be \emph{governed} also warrants the interest of social scientists.
However, it is not enough to show citizens' opinions on governance
proposals, but also to explore what factors influence these demands.
This is also of interest for the scholarly AI ethics community as
factors may be identified that contribute to critical engagement with AI
technology. Researchers already have explored a plethora of influential
factors in relation to AI \emph{opinions} and AI \emph{use}. However,
less attention has been paid to governance issues.

Therefore, in this study, we aim to contribute empirical evidence on
what factors lead citizens to demand stronger regulatory approaches. In particular, we focus
on the roles of discrimination perceptions of RBI technologies as well
as trust in AI technology and law enforcement as a user of RBI
technology. We do this with respect to the use of RBI in different usage
contexts: 1) use with the purpose of identifying criminals vs.~securing
public events (such as the Olympics), and 2) use of RBI after a criminal
activity vs.~in real time. To gather empirical data, we used a factorial
survey of \emph{n=983} German citizens.

\hypertarget{the-need-for-governance-of-ai}{%
\section{The Need for Governance of
AI}\label{the-need-for-governance-of-ai}}

The use of AI in the public sector is a particularly sensitive area of
application, as it can have an impact on the lives of citizens without
them having the choice of whether or not to engage with an AI system.
There are numerous examples where AI systems have led to
detrimental consequences for certain citizens \cite{AlgorithmWatch.2020}. For example, a
tax fraud detection system used by the tax authorities in the
Netherlands discriminated against people with certain demographic
characteristics, such as gender, age and place of residence \cite{Constantaras.2023, Heikkila.2022}, leading to the misclassification of thousands of people. These
people were ultimately denied access to social services, which had a
serious impact on their lives. In the UK, an automated grading system
was used to determine students' final exam grades. As most students
received significantly lower grades than before, this led to a public
protest that resulted in the abolition of the system \cite{Kelly.2021}. Also, in
the education sector, an ADM system used for university admissions in
France has been abandoned due to fairness issues arising from its use \cite{Wenzelburger.2022}. In the field of public security, there is a huge scientific
and political discussion about the use of AI in the criminal justice
sector, for example, for bail setting or predictive policing \cite{Angwin.2016, Hartmann.2021}.

All of these cases have in common that the AI system in question can be considered to have a high impact on the lives of citizens. However, the use of AI in the public sector is not profit-driven and should serve the public interest \cite{Zuger.2022}. Accordingly, the use of AI must be justified to the public and it must be ensured that the use of AI technology does not cause social harm, i.e., that it meets ethical standards \cite{Zuger.2022}. This approach is consistent with the normative goal of creating AI according to the idea of the Common Good \cite{Berendt.2019, Floridi.2018, Foffano.2022}, which aims to maximize ethical standards in AI that serve all affected individuals and not just a few stakeholders. However, ethical goals are often traded off against rapid implementation by vendors and developers, due to a strong focus on economic growth and a trial-and-error mentality \cite{Gerdes.2022}. This trend can be seen not only in the economic sector, but also in the political arena. The global AI race leads policymakers to prioritize funding and regulation for economic progress, which can lead to the neglect of ethical issues \cite{Cave.2018}. As a result, policymakers present AI as an inevitable technology that must be advanced quickly to keep pace with other nations \cite{Bareis.2021, Katzenbach.2021}. Following this approach, a lot of money is invested in the technology sector, while societal voices are largely ignored \cite{GaldonClavell.2023, Gerdes.2022}.

Nevertheless, AI ethics guidelines have been published by several entities, suggesting ways to mitigate the ethical risks of AI systems \cite{AlgorithmWatch.2023, Fjeld.2020, Franzke.2022, Hagendorff.2020, Jobin.2019, Roche.2022}. However, these guidelines lack a reinforcement mechanism \cite{Hagendorff.2020, Kieslich.2021c}. Stronger regulation could be such a mechanism, leading to the enforcement of ethical standards for AI. Several scholars in the field of AI ethics have already pointed to the need for a regulatory framework for AI \cite{Crawford.2021, Krafft.2022}. With the EU AI Act, such a framework has now been agreed upon and will be implemented in the EU member states. In principle, the EU AI Act distinguishes between four risk categories: unacceptable risk, high risk, limited risk, and minimal risk \cite{Barkane.2022, EuropeanParliament.2023}. While systems in the unacceptable risk category should be banned outright, systems in the other categories require different forms of regulation to limit the potential negative consequences.

\hypertarget{remote-biometric-identification-and-the-ai-act}{%
\section{Remote Biometric Identification and the AI
act}\label{remote-biometric-identification-and-the-ai-act}}

In this study, we are particularly interested in RBI systems. Law enforcement agencies can use RBI as a predictive policing tool in the pursuit of criminals. The technology enables both real-time identification of people and post hoc analysis of collected video footage. Most RBI systems use facial recognition technology and compare faces to a database \cite{Neuwirth.2023}.

The motivation to deploy RBI systems in the first place is mostly driven by security concerns, largely influenced by the threat of terrorism affecting society on a large scale \cite{Brown.2009}. There is a widespread belief among policy makers and law enforcement agencies that the use of RBI systems will help in the fight against crime \cite{vanBrakel.2011} and that the ``increased use of technology
[...] will render policing more
efficient, whether this statement is eventually proven to be true or
not.'' \cite{Vogiatzoglou.2019}. This changes the focus of law enforcement from prosecuting crimes that have already been committed to a strategy that attempts to prevent crime \cite{Vogiatzoglou.2019}. However, the use of RBI systems requires the collection of massive amounts of data that are needed to make predictions more reliable. In this sense, these systems are very privacy invasive \cite{Klitou.2014}, or, as Vogiatzoglou \cite{Vogiatzoglou.2019} describes it: ``Data are
gathered not for a specific criminal investigation but rather for an
undetermined purpose, serving a mentality of `nice-to-have' rather than
`must-have' intelligence. {[}\ldots{]} Serving this mentality of
`nice-to-have', practices of mass surveillance have increasingly become the
most popular means used by both law enforcement and intelligence
services in the fight against serious crime.''

Consequently, the large-scale application of RBI raises issues about the
violation of fundamental rights of citizens, for instance, in terms of
privacy, stereotyping, de-individualization and non-discrimination \cite{Barkane.2022, Neuwirth.2023, Vogiatzoglou.2019}. Furthermore, doubts exist concerning the accuracy,
reliability, and security of those systems \cite{Neuwirth.2023}. Acknowledging
these risks, the EU AI Act makes several distinctions in the
regulation of RBI. Regarding the \emph{temporal} aspect, real-time data
analysis should be \emph{prohibited} with the three exceptions of
searching for victims of crime and/or missing children, the prevention
of serious crimes such as terrorist attacks, and the prosecution of
criminal offenders \cite{Barkane.2022, Neuwirth.2023}. The post hoc analysis is \emph{allowed}
under legal obligations (e.g.~usage has to be authorized by a judge) \cite{Barkane.2022, EuropeanParliament.2023}. The risk classification of RBI systems under the EU AI Act is,
thus, dependent on the \emph{temporal} aspect of data analysis (post hoc
analysis vs.~real-time) and the \emph{purpose} of its usage. Real-time
RBI systems should be banned with some exceptions, whereas post hoc
analysis RBI systems are classified as high-risk systems. Those
high-risk systems, according to the EU AI Act, ``must implement a risk
management system, use high-quality data sets, draw up technical
documentation, enable record-keeping, ensure transparency and provide
information to users, ensure human oversight and an appropriate level of
robustness, accuracy and cybersecurity'' \cite{Barkane.2022}. Additionally, these
systems need to be audited by third parties and registered into an EU
database \cite{Barkane.2022}.

At the time of writing, the final draft of the EU AI Act has not yet been published, and there are reports of exceptions to the real-time use of RBI. In the run-up to the final discussion and vote on the EU AI Act, some political interest groups and parties have opted for more exemptions for the real-time use of RBI. In addition, some national governments also pushed for broader use of real-time RBI (e.g., the French government advocated the use of RBI at the 2024 Olympic Games) in the interest of national security \cite{Kayali.2023}. Many scholars have criticized that the exceptions made for the use of real-time RBI open loopholes for widespread use of the technology \cite{Barkane.2022, Neuwirth.2023, Veale.2021}.

\hypertarget{public-opinion-on-ai-governance}{%
\section{Public Opinion on AI
Governance}\label{public-opinion-on-ai-governance}}

Given the enormous consequences of the use of RBI for society, the question of society's influence on the regulation of AI remains open. According to Rahwan \cite{Rahwan.2018}, society needs to be involved in
setting norms and regulations on how society wants to interact with AI.
In his Society-in-the-Loop (SITL) approach, he opts to integrate
public opinion on moral and ethical decisions into the regulatory
framework. The call for greater public involvement is shared by several
other scholars. For example, Züger and Asghari \cite{Zuger.2022} argue for strengthening public
deliberation on AI issues to include the voice of all members of society, and Crawford \cite{Crawford.2021} opts for stronger counter-movements against
techno-deterministic approaches. All in all, the AI for the Common Good
and Public Interest community is united by the call for a stronger
inclusion of societal voices to counterbalance hegemonic political and
economic approaches \cite{Gerdes.2022b, Rudschies.2021, Sartori.2022, Yeung.2020, Zuger.2022}. Taking this call seriously
means exploring public opinion on pressing issues related to AI and
regulation. The potential use of RBI is arguably one such use case.

Public opinion can be a crucial factor in shaping technology development. In
democratic societies, the public can express its political engagement in
a variety of ways, such as protesting or, at the most basic level,
voting. However, public opinion can only exert pressure if an issue is
on the political agenda, i.e.~if it is perceived as relevant by the
public. The politicization of an issue is, therefore, a crucial factor in
public influence. An issue can be perceived as politicized when the
public debate is polarized, the issue is intensively reported on, and the
issue resonates in society \cite{Schattschneider.1957, Wilde.2011}. However, recent surveys from
the German context show that AI issues are not overly salient in the
German population -- in particular, AI ethics issues are not at all on
the agenda for most German citizens \cite{Kieslich.2023}. Moreover, most
German citizens do not consider AI issues overly relevant for
their general voting decisions \cite{MeinungsmonitorKunstlicheIntelligenz.2023}.

However, empirical studies focusing on \emph{specific use cases} have
shown that the public can be engaged with AI issues. For example, Marcinkowski and colleagues \cite{Marcinkowski.2020} showed that perceptions of unfairness lead to intentions to
protest against university admissions systems, and Lünich and Kieslich \cite{Lunich.2022} showed that
distrust of AI systems leads to them being perceived as illegitimate. As mentioned above, several counter-movements against some AI
applications have emerged \cite{Crawford.2021}. Further, perceptions of trust in AI
have been shown to be a critical factor in technology adoption \cite{Bareis.2021, Shin.2020, Shin.2021, Shin.2021b}. Additionally, Kieslich and colleagues \cite{Kieslich.2023} have shown that awareness of ethical
issues of AI leads to higher political engagement.

But it is not only AI-related attitudes that influence political
attitudes toward AI. Wenzelburger and colleagues \cite{Wenzelburger.2022} point out that contextual factors also
play a role. In an empirical study on AI adoption in the public sector
they showed that ``the personal importance of the problem that an
algorithm is supposed to deal with and the values at stake clearly
matter for the extent to which citizens show general support of
algorithms in policing'' \cite{Wenzelburger.2022}. In addition, they also report that
trust in the organisation using AI positively influences its acceptance,
while the technological performance of the system has only marginal
effects on technology acceptance \cite{Wenzelburger.2022}.

In addition, several studies have been conducted that focus on the acceptance of facial recognition technology. Ritchie and colleagues empirically assessed public attitudes toward automatic facial recognition technology in the United Kingdom, Australia, and the United States \cite{Ritchie.2021}. They report that the context of use affects support for the use of facial recognition technology in the criminal justice system. Kostka and colleagues examined public opinion on the acceptability of the use of facial recognition technology in China, Germany, the United Kingdom, and the United States \cite{Kostka.2023}. Their survey study distinguished four sets of influencing factors: political context and attitudes, history of surveillance, concerns about public issues, and individual preferences and characteristics. Their findings for the German context suggest that people who perceive facial recognition technology as a general invasion of privacy and a risk tend not to support the use of facial recognition technology. Moreover, German citizens concerned about terrorism and socially unacceptable public behavior also support the use of facial recognition technology. For all countries except Germany, effects were also found for trust in government and technology affinity. For example, people who trust the government and are open to technology also support the use of facial recognition technology. In addition, Trüdinger and Steckermeier showed that political trust leads to higher acceptance of surveillance policies in Germany \cite{Trudinger.2017}.

However, there is still a lack of research on citizens' demand for concrete governance strategies on AI; in particular, to our knowledge, no other study has focused on the link between public opinion on AI and concrete regulatory measures regarding the EU AI Act. In this study, we focus on three different policy proposals: we distinguish between different regulatory mechanisms in 1) banning the technology due to its classification as an unacceptably risky system, and 2) registering the technology in a public database, as well as 3) the need for independent third-party review, as required in the high-risk category of the EU AI Act \cite{Barkane.2022}. Furthermore, we contribute to the research literature by investigating different explanatory factors for regulatory demands: discrimination perceptions of RBI systems, trust in AI, and trust in law enforcement. We do this with respect to the use of RBI in different contexts: 1) use with the purpose of identifying criminals vs. securing public events, and 2) use of RBI after a criminal activity vs. in real time.

\hypertarget{research-questions}{%
\section{Research Questions}\label{research-questions}}

We first focus on whether the conditions of RBI deployment
matter in terms of the regulation demands of the citizens. As outlined
earlier, the EU AI Act makes several distinctions regarding the
regulation of RBI. We manipulate two conditions of the use of RBI: the
\emph{temporal} aspect (post hoc analysis vs.~real-time) and
\emph{purpose of use} (prosecuting criminals vs.~securing public
events). Thus, we pose the following research questions.

\emph{RQ1: Does the temporal context of data analysis (post hoc analysis
vs.~real-time) affect the approval of stronger regulatory interventions
regarding RBI?}

\emph{RQ2: Does the purpose of the use of RBI (prosecuting criminals
vs.~securing public events) affect the approval of stronger regulatory
interventions regarding RBI?}

Further, we explore the mediating role of discrimination perceptions as
a distinct measure of awareness of AI ethics. We ask the question
of whether discrimination perceptions are dependent on the context of the use of RBI. Accordingly, we ask:

\emph{RQ3: How do perceptions of discrimination mediate the relationship
between the temporal context and the support for stronger regulatory
interventions regarding RBI?}

In the context of AI, trust is often recognized as a critical factor,
for example, the EU's policy strategy is to develop a trustworthy
approach to AI \cite{EuropeanCommission.2019}, and trust in AI was also identified as a
critical factor for AI adoption \cite{Aysolmaz.2023, Lunich.2022, Shin.2020, Shin.2021, Shin.2021b}. In addition, trust
in law enforcement, i.e.~the agent using RBI systems, was identified as
a contributing factor to the use of technology in law enforcement \cite{Kostka.2023, Trudinger.2017}. Thus, we include trust in AI and trust in law enforcement as
independent variables in our model. We are interested in the direct
effect of the approval of stronger regulatory interventions as well as
on the mediation effects of perceptions of discrimination. Hence, we
ask:

\emph{RQ4a: Does trust in AI affect the approval of stronger regulatory
interventions regarding RBI?}

\emph{RQ4b: Does trust in law enforcement affect the approval of
stronger regulatory interventions regarding RBI?}

\emph{RQ5a: How do perceptions of discrimination mediate the
relationship between trust in AI and the support for stronger regulatory
interventions regarding RBI?}

\emph{RQ5b: How do perceptions of discrimination mediate the
relationship between trust in law enforcement and the support for
stronger regulatory interventions regarding RBI?}

In addition, we also include sociodemographic variables as controls, as
several studies have shown that these influence technology acceptance or
political engagement \cite{Carradore.2021, Fietta.2021, Ikkatai.2022, Kieslich.2022, Kieslich.2023, Zhang.2019}. Specifically, we pose
the following research questions:

\emph{RQ6: Do other contextual factors affect the approval of stronger
regulatory interventions regarding RBI?}

\emph{RQ7: How do perceptions of discrimination mediate the relationship
between the other contextual factors and the support for stronger
regulatory interventions regarding RBI?}

\hypertarget{procedure-sample}{%
\subsection{Procedure \& Sample}\label{procedure-sample}}

To answer our research questions, we conducted an online survey among
German citizens. The data was collected from June 19 to July 7, 2023. To
recruit participants, we used the SoSci Panel, which is based on a
convenience sample of German-speaking respondents. The SoSci Panel is a
joint project of the Institute for Communication Science in Munich and
the German Society for Communication Science (DGPuK). As such, it is
thoroughly maintained, has strict quality criteria, and an internal
peer-review process for studies conducted with the panel. However, the
results are not representative of the German population.

As our study aims to explore political engagement and support for
governance mechanisms, we decided to include only German citizens in the
final sample. The inclusion of other political contexts (such as Swiss
or Austrian), while interesting, would have added another layer of
complexity to this study that we could not satisfactorily address.

The survey was designed as follows: First, participants had to answer
some questions about their media use as well as their personal beliefs
about social and political issues. The term AI was then introduced with
the following definition: ``\emph{There is currently a lot of public
talk about `artificial intelligence' (AI). What is meant here are
computer applications that automatically evaluate digital data. For AI,
the evaluation of large volumes of data AI represents a learning process
from which it continuously processes new information and thus recognizes
ever more precise patterns over time. Based on this analysis, facts can
be determined, and future developments can be predicted.
Artificial intelligence systems can suggest courses of action or make
decisions autonomously and or make decisions autonomously and execute
them directly.}'' Immediately afterwards, the participants had to
indicate their trust in AI. We then introduced the use case of our study
-- Remote Biometric Identification. We manipulated the use of RBI systems
in terms of two factors: 1) temporal component (post hoc analysis
vs.~real-time analysis) and 2) purpose of use (identifying criminals
vs.~securing public events). Participants were exposed to one of four
possible scenarios. After the treatment check, participants answered
questions about their concerns regarding the discriminatory impact of
RBI. In addition, we asked our dependent variables about support for
government action on RBI. Finally, we collected sociodemographic
information and measured whether participants followed
the EU AI Act debate. The survey was conducted in German.

All in all, 1003 respondents participated in the survey. However, 20
cases had to be excluded as these participants failed the treatment
check. Thus, our final sample consists of 983 cases. 538 participants
identify as female, 434 identify as male and 11 identify as non-binary.
The average age of the respondents is 50 years (\emph{SD}=18.25).

\hypertarget{measures}{%
\subsection{Measures}\label{measures}}

\hypertarget{dependent-variables-support-of-regulatory-interventions}{%
\subsubsection{Dependent Variables: Support of Regulatory
Interventions}\label{dependent-variables-support-of-regulatory-interventions}}

We queried the support for regulatory interventions regarding AI with
three self-developed items on a five-point Likert scale (1=do not
support at all; 5=totally support; -1=don't know). All regulatory
interventions are derived from current AI regulations that are either
proposed by the EU AI Act or scholars in the field \cite{EuropeanCommission.2021, Shneiderman.2022b}. The
question and items read as follows: \emph{At the political level, the
regulation of AI-based remote biometric identification systems is
currently being discussed. How much do you agree that policy should set
the following rules?} Remote biometric identification should be banned;
Law enforcement agencies should be required to commission independent
testing companies (e.g.~TÜV\footnote{TÜV (short for Technischer
  Überwachungsverein {[}technical inspection association{]}) are German
  oversight organizations for technical safety checks. The TÜV is
  well-known in Germany as they are also conducting mandatory car
  inspections.}) to test the dangers of the system before introducing
remote biometric identification; Law enforcement agencies should be
required to register remote biometric identification systems and their
modes of operation in a public database. The three items will
subsequently be addressed in short form as (1) ban, (2) audit, and (3)
database registration.

\hypertarget{mediator-discrimination-perception}{%
\subsubsection{Mediator: Discrimination
Perception}\label{mediator-discrimination-perception}}

Discrimination perception through
RBI was measured via five items on a five-point Likert scale (1=do not
support at all; 5=totally support; -1=don't know) that were adapted from Kieslich et al. \cite{Kieslich.2020}. The items that show good factorial validity and internal 
consistency (Cronbach's \(\alpha\) = 0.88, Average Variance Extracted
(AVE) = 0.60) read as follows: \emph{If you now think about the
consequences of using AI for remote biometric identification. To what
extent do you agree or disagree with the following statements?}  The use of RBI creates injustices. RBI
systematically puts certain groups of people at a disadvantage. Existing
inequalities are reinforced by the use of RBI. RBI creates new
inequalities. The use of RBI leads to discrimination.

\hypertarget{independent-variables}{%
\subsubsection{Independent Variables}\label{independent-variables}}

\textbf{Use case of RBI.} As outlined above, respondents were confronted
with one of four possible stimuli. We manipulated the \emph{temporal}
aspect (0 = post hoc analysis, 1 = real-time analysis) and the
\emph{purpose of use} (0 = public event, 1 = prosecution of criminals).
The wording of the stimuli can be found in the Appendix.

\textbf{Trust in AI.} For trust in AI, respondents rated four statements
on a five-point Likert scale (1=do not support at all; 5=totally
support). The scale was adopted from Lünich and Kieslich \cite{Lunich.2022} as well as Shin \cite{Shin.2021}. The
items that show good factorial validity and internal consistency
(Cronbach's \(\alpha\) = 0.90, AVE = 0.69) read as follows: I trust that
AI systems can make correct decisions; I trust the decisions made by AI
systems; Decisions made by AI systems are trustworthy; I believe that
decisions made by AI systems are reliable.

\textbf{Trust in Police and Law Enforcement.} Trust in police and law
enforcement was measured with the same item wording and scale as the trust
in AI except replacing the word ``AI'' with ``police and law
enforcement''. Again, this scale was adopted from  Lünich and Kieslich \cite{Lunich.2022} and Shin \cite{Shin.2021}.
The items that show good factorial validity and internal consistency
(Cronbach's \(\alpha\) = 0.96, AVE = 0.86) read as follows: \emph{To
what extent do you agree or disagree with the following statements?} I
trust that German police and law enforcement authorities can make the
right decisions; I trust the decisions made by German police and law
enforcement authorities; Decisions made by German police and law
enforcement authorities are trustworthy; I believe that the decisions
made by German police and law enforcement authorities are reliable.

\hypertarget{controls}{%
\subsubsection{Controls}\label{controls}}

\textbf{Experienced Discrimination.} Experienced discrimination was
measured on a five-point Likert scale (1=never; 2=seldom; 3=sometimes;
4=often; 5=very often). The item formulation was adopted from Schumann and colleagues \cite{Schumann.2019},
which is the German translation of the scale from Williams et al. \cite{Williams.1997}. The items
that show good factorial validity and internal consistency (Cronbach's
\(\alpha\) = 0.88, AVE = 0.60) read as follows: \emph{How often have any
of the following things happened to you in your everyday life?} You were
treated with less respect than other people; Someone acted as if you
were not taken seriously; You were threatened or harassed.

\textbf{Domestic Security Concerns.} Domestic Security Concerns was measured with one
item on a four-point Likert scale (1=not at all concerned; 2=a little
concerned; 3=concerned; 4=very concerned) that was adopted from Wenzelburger et al. \cite{Wenzelburger.2022}. It reads: \emph{Please indicate - intuitively - how worried
you are about internal security in Germany.}

\textbf{Other Controls.} Additionally, we queried if the respondents
have heard about the EU AI Act before (1=yes), as well as gender gender
(1=female) and age (in years).

\hypertarget{results}{%
\section{Results}\label{results}}

We conducted our data analysis using a structural regression model,
employing the \emph{lavaan} package in R \cite{Rosseel.2012}. The model
encapsulated both the direct effects of the stimuli and other
independent variables on the three dependent variables concerning
support for regulatory interventions, and the indirect effects mediated
through perceptions of resulting discrimination by RBI.

We subsequently present the results for each dependent variable separately when focusing on direct and indirect effects for enhanced clarity and comprehension. Table 1 below details the support for a ban on RBI, Table 2 presents findings on mandatory auditing of RBI systems, and
Table 3 focuses on the necessity for registering all RBI systems in a
public database. Given that our convenience sample size was not
determined by a priori power analysis, we will subsequently concentrate
on results where total standardized effects exceed .1 as the smallest
effect size of interest. This approach will ensure a more focused
examination of practically significant findings, emphasizing effects
that are not only statistically significant but also of substantive
importance in the context of our study. Overall, the structural
regression model shows good fit (\(\chi^2\)(224) = 605.03, \emph{p} < 0.001; \emph{RMSEA} = 0.042, \emph{90\% CI} {[}0.038, 0.046{]};
\emph{TLI} = 0.963).

\hypertarget{banning-rbi}{%
\subsection{Banning RBI}\label{banning-rbi}}

Concerning the policy advocating for a complete ban on RBI (see Table
1), the findings for the total effects indicate no significant
differences attributable to either the temporal condition (B = 0.010,
\emph{SE} = 0.094, \emph{95\% CI} (-0.174, 0.195), \emph{p} = 0.913,
\(\beta\) = 0.003) or the purpose of use (B = -0.035, \emph{SE} = 0.094,
\emph{95\% CI} (-0.219, 0.150), \emph{p} = 0.712, \(\beta\) = -0.011).

Regarding the direct effects, on the one hand, the results reveal that
both a higher trust in law enforcement (B = -0.236, \emph{SE} = 0.054,
\emph{95\% CI} (-0.341, -0.131), \emph{p} < 0.001, \(\beta\) = -0.140)
and a greater trust in AI among citizens (B = -0.215, \emph{SE} = 0.059,
\emph{95\% CI} (-0.330, -0.099), \emph{p} < 0.001, \(\beta\) = -0.110)
are linked to decreased support for a ban on RBI. Moreover, results
suggest that regarding the support for a ban on RBI, discrimination
perceptions had a strong positive effect (B = 0.543, \emph{SE} = 0.041,
\emph{95\% CI} (0.462, 0.624), \emph{p} < 0.001, \(\beta\) = 0.413). As
a consequence, discrimination perceptions mediate these effects of trust
in law enforcement (B = -0.190, \emph{SE} = 0.028, \emph{95\% CI}
(-0.244, -0.136), \emph{p} < 0.001, \(\beta\) = -0.113) and trust in AI
(B = -0.133, \emph{SE} = 0.028, \emph{95\% CI} (-0.188, -0.077),
\emph{p} < 0.001, \(\beta\) = -0.068) on supporting a ban on RBI as
suggested by the indirect effects, underscoring the influence of trust
levels on attitudes towards RBI regulation. Moreover, when citizens had
previously heard of the AI Act, there was a direct positive effect on
support for a RBI ban (B = 0.269, \emph{SE} = 0.106, \emph{95\% CI}
(0.061, 0.477), \emph{p} = 0.011, \(\beta\) = 0.071). This effect was
mediated by discrimination perceptions (B = 0.165, \emph{SE} = 0.050,
\emph{95\% CI} (0.067, 0.263), \emph{p} < 0.001, \(\beta\) = 0.044).
Awareness of the AI Act came with heightened perceptions of
discrimination, which in turn contributed to increased support for the
ban.

\begin{table*}
\centering
\resizebox{\linewidth}{!}{
\begin{tabular}{llllllll}
\toprule
\multicolumn{5}{c}{ } & \multicolumn{2}{c}{\textbf{95\% Confidence Interval}} & \multicolumn{1}{c}{ } \\
\cmidrule(l{3pt}r{3pt}){6-7}
 & \textbf{Estimate} & \textbf{Std. Error} & \textbf{z-value} & \textbf{p-value} & \textbf{Lower} & \textbf{Upper} & \textbf{Std. All}\\
\midrule
\textbf{Direct Effects} &  &  &  &  &  &  & \\
Temporal Condition (0 = post hoc analysis, 1 = real-time analysis -> Ban & -0.006 & 0.087 & -0.067 & 0.946 & -0.176 & 0.164 & -0.002\\
Purpose of Use (0 = Public Event, 1 = Prosecution of Criminals -> Ban & -0.011 & 0.087 & -0.129 & 0.898 & -0.181 & 0.159 & -0.004\\
Trust in Law Enforcement -> Ban & -0.236 & 0.054 & -4.400 & <.001 & -0.341 & -0.131 & -0.140\\
Trust in AI -> Ban & -0.215 & 0.059 & -3.644 & <.001 & -0.330 & -0.099 & -0.110\\
Experienced Discrimination -> Ban & -0.181 & 0.082 & -2.213 & 0.027 & -0.341 & -0.021 & -0.077\\
'Heard of AI Act?' (1 = Yes) -> Ban & 0.269 & 0.106 & 2.539 & 0.011 & 0.061 & 0.477 & 0.071\\
Domestic Security Concerns -> Ban & -0.051 & 0.055 & -0.928 & 0.353 & -0.159 & 0.057 & -0.027\\
Gender (1 = female) -> Ban & -0.207 & 0.090 & -2.316 & 0.021 & -0.383 & -0.032 & -0.066\\
Age -> Ban & -0.003 & 0.003 & -1.201 & 0.230 & -0.008 & 0.002 & -0.037\\
Mediator: Discrimination Perceptions -> Ban & 0.543 & 0.041 & 13.097 & <.001 & 0.462 & 0.624 & 0.413\\
\textbf{Indirect Effects} &  &  &  &  &  &  & \\
Temporal Condition (0 = post hoc analysis, 1 = real-time analysis -> Discrimination Perceptions -> Ban & 0.016 & 0.040 & 0.405 & 0.686 & -0.062 & 0.094 & 0.005\\
Purpose of Use (0 = Public Event, 1 = Prosecution of Criminals -> Discrimination Perceptions -> Ban & -0.024 & 0.040 & -0.590 & 0.555 & -0.102 & 0.055 & -0.007\\
Trust in Law Enforcement -> Discrimination Perceptions -> Ban & -0.190 & 0.028 & -6.871 & <.001 & -0.244 & -0.136 & -0.113\\
Trust in AI -> Discrimination Perceptions -> Ban & -0.133 & 0.028 & -4.664 & <.001 & -0.188 & -0.077 & -0.068\\
Experienced Discrimination -> Discrimination Perceptions -> Ban & 0.060 & 0.038 & 1.573 & 0.116 & -0.015 & 0.134 & 0.025\\
'Heard of AI Act?' (1 = Yes) -> Discrimination Perceptions -> Ban & 0.165 & 0.050 & 3.307 & <.001 & 0.067 & 0.263 & 0.044\\
Domestic Security Concerns -> Discrimination Perceptions -> Ban & -0.092 & 0.026 & -3.515 & <.001 & -0.143 & -0.041 & -0.049\\
Gender (1 = female) -> Discrimination Perceptions -> Ban & 0.030 & 0.041 & 0.735 & 0.462 & -0.050 & 0.111 & 0.010\\
Age -> Discrimination Perceptions -> Ban & -0.001 & 0.001 & -1.166 & 0.243 & -0.004 & 0.001 & -0.017\\
\textbf{Total Effects} &  &  &  &  &  &  & \\
Temporal Condition (0 = post hoc analysis, 1 = real-time analysis -> Ban & 0.010 & 0.094 & 0.110 & 0.913 & -0.174 & 0.195 & 0.003\\
Purpose of Use (0 = Public Event, 1 = Prosecution of Criminals -> Ban & -0.035 & 0.094 & -0.369 & 0.712 & -0.219 & 0.150 & -0.011\\
Trust in Law Enforcement -> Ban & -0.425 & 0.056 & -7.599 & <.001 & -0.535 & -0.316 & -0.253\\
Trust in AI -> Ban & -0.347 & 0.063 & -5.519 & <.001 & -0.471 & -0.224 & -0.178\\
Experienced Discrimination -> Ban & -0.121 & 0.088 & -1.368 & 0.171 & -0.294 & 0.052 & -0.051\\
'Heard of AI Act?' (1 = Yes) -> Ban & 0.434 & 0.115 & 3.791 & <.001 & 0.210 & 0.659 & 0.115\\
Domestic Security Concerns -> Ban & -0.143 & 0.059 & -2.407 & 0.016 & -0.259 & -0.027 & -0.076\\
Gender (1 = female) -> Ban & -0.177 & 0.097 & -1.821 & 0.069 & -0.368 & 0.014 & -0.056\\
Age -> Ban & -0.005 & 0.003 & -1.601 & 0.109 & -0.010 & 0.001 & -0.053\\
\bottomrule
\end{tabular}}
\caption{\label{ban}Mediation model on support of banning RBI}
\end{table*}

\hypertarget{auditing-rbi-systems}{%
\subsection{Auditing RBI systems}\label{auditing-rbi-systems}}

Concerning the demand for mandatory audits of RBI systems (see Table 2),
the findings for the total effects also indicate no significant
differences attributable to either the temporal condition (B = -0.140,
\emph{SE} = 0.099, \emph{95\% CI} (-0.333, 0.054), \emph{p} = 0.157,
\(\beta\) = -0.045) or the purpose of use (B = 0.175, \emph{SE} = 0.098,
\emph{95\% CI} (-0.018, 0.368), \emph{p} = 0.075, \(\beta\) = 0.056).

Moreover, neither trust in law enforcement (B = -0.032, \emph{SE} =
0.059, \emph{95\% CI} (-0.147, 0.083), \emph{p} = 0.587, \(\beta\) =
-0.019) nor trust in AI had a significant total effect on the demand for
mandatory audits of RBI systems (B = -0.069, \emph{SE} = 0.066,
\emph{95\% CI} (-0.198, 0.061), \emph{p} = 0.298, \(\beta\) = -0.036).

When it comes to significant direct effects, respondents' age was the
only factor showing a negative direct effect on support for mandatory
RBI audits (B = -0.009, \emph{SE} = 0.003, \emph{95\% CI} (-0.015,
-0.003), \emph{p} = 0.002, \(\beta\) = -0.109). Specifically, older
respondents were less inclined to favour these audits. However,
discrimination perceptions whose effect on the approval of mandatory
auditing of RBI systems was less pronounced regarding audit demands than
it was for a proposed ban (B = 0.162, \emph{SE} = 0.047, \emph{95\% CI}
(0.069, 0.254), \emph{p} < 0.001, \(\beta\) = 0.124) did not mediate the
effect of age on support for mandatory RBI audits (B = 0.000, \emph{SE}
= 0.000, \emph{95\% CI} (-0.001, 0.000), \emph{p} = 0.268, \(\beta\) =
-0.005).

\begin{table*}
\centering
\resizebox{\linewidth}{!}{
\begin{tabular}{lrrrlrrr}
\toprule
\multicolumn{5}{c}{ } & \multicolumn{2}{c}{\textbf{95\% Confidence Interval}} & \multicolumn{1}{c}{ } \\
\cmidrule(l{3pt}r{3pt}){6-7}
 & \textbf{Estimate} & \textbf{Std. Error} & \textbf{z-value} & \textbf{p-value} & \textbf{Lower} & \textbf{Upper} & \textbf{Std. All}\\
\midrule
\textbf{Direct Effects} &  &  &  &  &  &  & \\
Temporal Condition (0 = post hoc analysis, 1 = real-time analysis -> Audit & -0.145 & 0.098 & -1.474 & 0.140 & -0.337 & 0.048 & -0.046\\
Purpose of Use (0 = Public Event, 1 = Prosecution of Criminals -> Audit & 0.182 & 0.098 & 1.861 & 0.063 & -0.010 & 0.374 & 0.058\\
Trust in Law Enforcement -> Audit & 0.025 & 0.061 & 0.408 & 0.683 & -0.094 & 0.143 & 0.015\\
Trust in AI -> Audit & -0.029 & 0.067 & -0.437 & 0.662 & -0.160 & 0.101 & -0.015\\
Experienced Discrimination -> Audit & -0.069 & 0.092 & -0.746 & 0.456 & -0.249 & 0.112 & -0.030\\
'Heard of AI Act?' (1 = Yes) -> Audit & 0.164 & 0.120 & 1.368 & 0.171 & -0.071 & 0.399 & 0.044\\
Domestic Security Concerns -> Audit & 0.083 & 0.062 & 1.329 & 0.184 & -0.039 & 0.205 & 0.045\\
Gender (1 = female) -> Audit & -0.015 & 0.101 & -0.146 & 0.884 & -0.213 & 0.184 & -0.005\\
Age -> Audit & -0.009 & 0.003 & -3.086 & 0.002 & -0.015 & -0.003 & -0.109\\
Mediator: Discrimination Perceptions -> Audit & 0.162 & 0.047 & 3.436 & <.001 & 0.069 & 0.254 & 0.124\\
\textbf{Indirect Effects} &  &  &  &  &  &  & \\
Temporal Condition (0 = post hoc analysis, 1 = real-time analysis -> Discrimination Perceptions -> Audit & 0.005 & 0.012 & 0.402 & 0.688 & -0.019 & 0.028 & 0.002\\
Purpose of Use (0 = Public Event, 1 = Prosecution of Criminals -> Discrimination Perceptions -> Audit & -0.007 & 0.012 & -0.582 & 0.560 & -0.031 & 0.017 & -0.002\\
Trust in Law Enforcement -> Discrimination Perceptions -> Audit & -0.056 & 0.018 & -3.155 & 0.002 & -0.092 & -0.021 & -0.034\\
Trust in AI -> Discrimination Perceptions -> Audit & -0.039 & 0.014 & -2.828 & 0.005 & -0.067 & -0.012 & -0.020\\
Experienced Discrimination -> Discrimination Perceptions -> Audit & 0.018 & 0.012 & 1.438 & 0.150 & -0.006 & 0.042 & 0.008\\
'Heard of AI Act?' (1 = Yes) -> Discrimination Perceptions -> Audit & 0.049 & 0.020 & 2.421 & 0.015 & 0.009 & 0.089 & 0.013\\
Domestic Security Concerns -> Discrimination Perceptions -> Audit & -0.027 & 0.011 & -2.491 & 0.013 & -0.049 & -0.006 & -0.015\\
Gender (1 = female) -> Discrimination Perceptions -> Audit & 0.009 & 0.013 & 0.719 & 0.472 & -0.016 & 0.034 & 0.003\\
Age -> Discrimination Perceptions -> Audit & 0.000 & 0.000 & -1.109 & 0.268 & -0.001 & 0.000 & -0.005\\
\textbf{Total Effects} &  &  &  &  &  &  & \\
Temporal Condition (0 = post hoc analysis, 1 = real-time analysis -> Audit & -0.140 & 0.099 & -1.417 & 0.157 & -0.333 & 0.054 & -0.045\\
Purpose of Use (0 = Public Event, 1 = Prosecution of Criminals -> Audit & 0.175 & 0.098 & 1.779 & 0.075 & -0.018 & 0.368 & 0.056\\
Trust in Law Enforcement -> Audit & -0.032 & 0.059 & -0.543 & 0.587 & -0.147 & 0.083 & -0.019\\
Trust in AI -> Audit & -0.069 & 0.066 & -1.041 & 0.298 & -0.198 & 0.061 & -0.036\\
Experienced Discrimination -> Audit & -0.051 & 0.093 & -0.550 & 0.582 & -0.232 & 0.130 & -0.022\\
'Heard of AI Act?' (1 = Yes) -> Audit & 0.213 & 0.120 & 1.779 & 0.075 & -0.022 & 0.448 & 0.057\\
Domestic Security Concerns -> Audit & 0.056 & 0.062 & 0.893 & 0.372 & -0.066 & 0.177 & 0.030\\
Gender (1 = female) -> Audit & -0.006 & 0.102 & -0.057 & 0.955 & -0.205 & 0.194 & -0.002\\
Age -> Audit & -0.010 & 0.003 & -3.211 & 0.001 & -0.016 & -0.004 & -0.113\\
\bottomrule
\end{tabular}}
\caption{\label{audit}Mediation model on support of stronger auditing of RBI}
\end{table*}

\hypertarget{database-registration-of-rbi-systems}{%
\subsection{Database registration of RBI
systems}\label{database-registration-of-rbi-systems}}

Concerning the demand for mandatory registration of RBI systems in a
public database (see Table 3), the findings for the total effects also
indicate no significant differences attributable to either the temporal
condition (B = 0.050, \emph{SE} = 0.130, \emph{95\% CI} (-0.205, 0.304),
\emph{p} = 0.703, \(\beta\) = 0.012) or the purpose of use (B = -0.156,
\emph{SE} = 0.129, \emph{95\% CI} (-0.409, 0.098), \emph{p} = 0.229,
\(\beta\) = -0.038).

Moreover, again, neither trust in law enforcement (B = -0.131, \emph{SE}
= 0.077, \emph{95\% CI} (-0.282, 0.020), \emph{p} = 0.089, \(\beta\) =
-0.060) nor trust in AI had a significant total effect on the demand for
mandatory database registration of RBI systems (B = -0.050, \emph{SE} =
0.087, \emph{95\% CI} (-0.220, 0.119), \emph{p} = 0.561, \(\beta\) =
-0.020).

Regarding the significant direct effects, respondents' gender was the
only factor showing a direct effect on support for mandatory
registration of RBI systems in a public database (B = -0.477, \emph{SE}
= 0.133, \emph{95\% CI} (-0.737, -0.217), \emph{p} < 0.001, \(\beta\) =
-0.116). Specifically, respondents who identified as female showed less
inclination towards favouring the registration of databases. However,
discrimination perceptions whose effect on the approval of mandatory
public database registration of RBI systems was again less pronounced
than it was for a proposed ban (B = 0.258, \emph{SE} = 0.062, \emph{95\%
CI} (0.137, 0.379), \emph{p} < 0.001, \(\beta\) = 0.151) did not mediate
the effect of gender on support for mandatory registration of RBI
systems in a public database (B = 0.014, \emph{SE} = 0.020, \emph{95\%
CI} (-0.025, 0.053), \emph{p} = 0.469, \(\beta\) = 0.003).

\begin{table*}
\centering
\resizebox{\linewidth}{!}{
\begin{tabular}{lrrrlrrr}
\toprule
\multicolumn{5}{c}{ } & \multicolumn{2}{c}{\textbf{95\% Confidence Interval}} & \multicolumn{1}{c}{ } \\
\cmidrule(l{3pt}r{3pt}){6-7}
 & \textbf{Estimate} & \textbf{Std. Error} & \textbf{z-value} & \textbf{p-value} & \textbf{Lower} & \textbf{Upper} & \textbf{Std. All}\\
\midrule
\textbf{Direct Effects} &  &  &  &  &  &  & \\
Temporal Condition (0 = post hoc analysis, 1 = real-time analysis -> Database Registration & 0.042 & 0.129 & 0.326 & 0.745 & -0.210 & 0.294 & 0.010\\
Purpose of Use (0 = Public Event, 1 = Terrorism -> Database Registration & -0.144 & 0.128 & -1.125 & 0.260 & -0.396 & 0.107 & -0.035\\
Trust in Law Enforcement -> Database Registration & -0.041 & 0.079 & -0.517 & 0.605 & -0.197 & 0.115 & -0.019\\
Trust in AI -> Database Registration & 0.013 & 0.087 & 0.145 & 0.884 & -0.159 & 0.184 & 0.005\\
Experienced Discrimination -> Database Registration & -0.035 & 0.121 & -0.291 & 0.771 & -0.272 & 0.202 & -0.011\\
'Heard of AI Act?' (1 = Yes) -> Database Registration & 0.086 & 0.157 & 0.545 & 0.586 & -0.223 & 0.394 & 0.017\\
Domestic Security Concerns -> Database Registration & -0.102 & 0.082 & -1.248 & 0.212 & -0.262 & 0.058 & -0.042\\
Gender (1 = female) -> Database Registration & -0.477 & 0.133 & -3.592 & <.001 & -0.737 & -0.217 & -0.116\\
Age -> Database Registration & 0.004 & 0.004 & 1.123 & 0.261 & -0.003 & 0.012 & 0.039\\
Mediator: Discrimination Perceptions -> Database Registration & 0.258 & 0.062 & 4.186 & <.001 & 0.137 & 0.379 & 0.151\\
\textbf{Indirect Effects} &  &  &  &  &  &  & \\
Temporal Condition (0 = post hoc analysis, 1 = real-time analysis -> Discrimination Perceptions -> Database Registration & 0.008 & 0.019 & 0.403 & 0.687 & -0.030 & 0.045 & 0.002\\
Purpose of Use (0 = Public Event, 1 = Terrorism -> Discrimination Perceptions -> Database Registration & -0.011 & 0.019 & -0.585 & 0.559 & -0.049 & 0.026 & -0.003\\
Trust in Law Enforcement -> Discrimination Perceptions -> Database Registration & -0.090 & 0.024 & -3.702 & <.001 & -0.138 & -0.042 & -0.041\\
Trust in AI -> Discrimination Perceptions -> Database Registration & -0.063 & 0.020 & -3.195 & 0.001 & -0.102 & -0.024 & -0.025\\
Experienced Discrimination -> Discrimination Perceptions -> Database Registration & 0.028 & 0.019 & 1.484 & 0.138 & -0.009 & 0.066 & 0.009\\
'Heard of AI Act?' (1 = Yes) -> Discrimination Perceptions -> Database Registration & 0.078 & 0.030 & 2.645 & 0.008 & 0.020 & 0.136 & 0.016\\
Domestic Security Concerns -> Discrimination Perceptions -> Database Registration & -0.044 & 0.016 & -2.743 & 0.006 & -0.075 & -0.012 & -0.018\\
Gender (1 = female) -> Discrimination Perceptions -> Database Registration & 0.014 & 0.020 & 0.725 & 0.469 & -0.025 & 0.053 & 0.003\\
Age -> Discrimination Perceptions -> Database Registration & -0.001 & 0.001 & -1.127 & 0.260 & -0.002 & 0.001 & -0.006\\
\textbf{Total Effects} &  &  &  &  &  &  & \\
Temporal Condition (0 = post hoc analysis, 1 = real-time analysis -> Database Registration & 0.050 & 0.130 & 0.382 & 0.703 & -0.205 & 0.304 & 0.012\\
Purpose of Use (0 = Public Event, 1 = Terrorism -> Database Registration & -0.156 & 0.129 & -1.202 & 0.229 & -0.409 & 0.098 & -0.038\\
Trust in Law Enforcement -> Database Registration & -0.131 & 0.077 & -1.703 & 0.089 & -0.282 & 0.020 & -0.060\\
Trust in AI -> Database Registration & -0.050 & 0.087 & -0.581 & 0.561 & -0.220 & 0.119 & -0.020\\
Experienced Discrimination -> Database Registration & -0.007 & 0.122 & -0.056 & 0.956 & -0.245 & 0.232 & -0.002\\
'Heard of AI Act?' (1 = Yes) -> Database Registration & 0.164 & 0.158 & 1.041 & 0.298 & -0.145 & 0.473 & 0.033\\
Domestic Security Concerns -> Database Registration & -0.146 & 0.082 & -1.782 & 0.075 & -0.306 & 0.015 & -0.060\\
Gender (1 = female) -> Database Registration & -0.462 & 0.134 & -3.453 & <.001 & -0.725 & -0.200 & -0.112\\
Age -> Database Registration & 0.004 & 0.004 & 0.943 & 0.346 & -0.004 & 0.012 & 0.033\\
\bottomrule
\end{tabular}}
\caption{\label{audit}Mediation model on support of registration of RBI in public database}
\end{table*}

\hypertarget{discussion}{%
\section{Discussion}\label{discussion}}

Our results yield important insights for the study of public opinion concerning
the different contextual use conditions of RBI, trust in AI, trust in
law enforcement, discrimination perceptions regarding AI technology, and
their practical relevance for implementing regulations.

\hypertarget{the-perceived-context-independence-of-rbi-use}{%
\subsection{The Perceived Context-Independence of RBI
Use}\label{the-perceived-context-independence-of-rbi-use}}

The absent effects of the experimental conditions -- both as direct and
mediated effects via discrimination perceptions -- suggest that differences concerning the specific design and aim of RBI systems are rather irrelevant to
citizens' demands for regulation. This is especially interesting in
consideration of the different risk classifications of the temporal
component in the EU AI Act \cite{EuropeanParliament.2023}. While the EU AI Act defines
real-time systems as unacceptably risky, its usage in post hoc analysis
is merely defined as high-risk. This difference is, however, not
mirrored in citizens' perceptions regarding regulatory demands.

However, it is unclear whether citizens are unable to distinguish between
the risk levels or whether they perceive them as equally risky. This
raises the question of whether citizens are overly concerned with how
these surveillance systems are ultimately designed. This may suggest a
restriction on the influence of public opinion in nuanced policy
matters. For example, respondents may not fully grasp the privacy
concerns associated with the real-time application of RBI. In such
scenarios, RBI surveillance occurs indiscriminately, monitoring all
citizens irrespective of any criminal activity. Conversely, in post hoc
analysis, the utilization of RBI is typically predicated on the
occurrence of a significant crime, warranting the examination of video
footage after the event. Nonetheless, considering the limited public
engagement with AI technology in general \cite{Kieslich.2023}, it is plausible that
many citizens are not adequately informed about the varying degrees of
impact associated with these technologies.

Consequently, political decision-makers and interest groups have even
more responsibility in recognizing the scope of different contexts of
the use of RBI. Taking public interest orientation seriously also
encompasses balancing the risks and benefits of AI systems, even if citizens
may not be aware of it. Given these insights, advocacy groups
championing marginalized communities or human rights should prioritize
conveying the potential risks associated with distinct RBI systems. They
ought to emphasize the privacy and discrimination concerns that can
arise from employing RBI, particularly in real-time scenarios.
Concurrently, law enforcement entities are likely to advocate for the
conditional use of RBI in exceptional circumstances, such as child
protection or counter-terrorism efforts. In the absence of a widespread
public debate, these nuanced yet critical discussions risk being
monopolized by specific stakeholder factions, excluding broader citizen
input. Consequently, the dynamics of power within these stakeholder
groups are poised to predominantly influence the ultimate decisions
regarding RBI usage, thereby reinforcing a top-down approach in AI
governance.

\hypertarget{trust-as-a-double-edged-sword-in-regard-to-strong-regulation}{%
\subsection{Trust as a Double-Edged Sword in Regard to Strong
Regulation}\label{trust-as-a-double-edged-sword-in-regard-to-strong-regulation}}

We found that trust plays a pivotal role in the support of banning RBI
technology. Considering the total effects of trust in law enforcement
and trust in AI on the dependent variables, we identified negative
effects for the demands for a ban, while we found no significant total
effects for the demand for registering RBIs in a public database and
their mandatory audit by third parties. If people trust in the
deployers and developers of the technology, they are less likely
demanding a ban on RBI. Trust, as such, can be interpreted as a passing
of responsibility towards other actors: either law enforcement, who
applies the technology, or the technology (or developing) companies,
that guarantee the functioning of the technology.

In addition, discrimination perceptions of RBI technology mediate the
effect of the trust variables on a demand for a ban. Trust in law
enforcement and trust in AI leads to lower perceptions of
the discriminatory impact of AI. If people trust the technology or
law enforcement, it also weakens discrimination perceptions, which,
otherwise, have a high positive impact on citizens' demand for a ban.
Thus, higher trust dampens this effect. Eventually, if the technology seems
trustworthy and citizens believe in the actors that apply the
technology, discrimination concerns diminish, and consequently, people
are less likely to opt for a ban on RBI technology.

These findings are in line with previous research focusing on the link
between trust in AI and AI adoption \cite{Aysolmaz.2023, Lunich.2022, Shin.2020, Shin.2021, Shin.2021b}. Trust leads
citizens to use and accept AI technology. However, several scholars also
noted the potential detrimental effects of overtrust, i.e.~trusting even
malfunctioning systems that may negatively affect the user and/or
society \cite{Krugel.2022, Robinette.2016}. Ultimately, it also depends on the system's technical
capabilities and societal effects if trusting AI systems
leads to positive outcomes in the sense of the public interests. In the
case of RBI systems, several scholars have warned regarding their
potential for human rights violations and their actual performance \cite{Barkane.2022, Neuwirth.2023, Vogiatzoglou.2019}. On the other side, law enforcement stresses the benefits for
security that these systems might entail \cite{Brown.2009, vanBrakel.2011}. Thus, in the end,
supporting regulatory governance policies is also a result of a
trade-off of perceptions between the benefits and risks of the technology.

These findings yield important implications for stakeholders who
communicate about AI. On the one hand, fueling distrust in AI may lead
to more awareness regarding discriminatory impacts, which in turn
influences the demand for strong regulations. On the other hand, the EU
strives for the development of trustworthy AI systems \cite{EuropeanCommission.2019}. That
results in the question of how and if a way can be found to do both --
strengthening critical evaluation and trust in (good) AI systems. Trust,
as found in our study, is more a double-edged sword. Consequently,
trust needs to be calibrated to ensure that citizens do not overtrust a
risky technology, which may lead to detrimental consequences for society.

\hypertarget{the-impact-of-additional-factors}{%
\subsection{The Impact of Additional
Factors}\label{the-impact-of-additional-factors}}

In addition to the effects of trust, we examined other factors that have
been found to steer public opinion towards AI and the
respondents' awareness of the AI Act itself. We found that having heard
of the AI Act impacts the demand for a ban on RBI technology. As the AI
Act focuses on risk classification -- and RBI are deemed as either
unacceptable or high risk -- it is plausible that the use of RBI in
the context of the AI Act is also perceived as risky and is connected to
discrimination perceptions as some interest groups are actively
highlighting exactly those harms. Therefore, individuals previously aware of the EU AI Act are also more likely to be informed about its ethical implications. Conversely, familiarity with the EU AI Act might indicate prior engagement with AI issues, indicating increased involvement in individual research, media coverage, or personal discussions related to the AI Act. This engagement in discussions about AI and its societal impact influences people's perceptions. While the exact causal relationship requires further investigation, this finding underscores the importance of involving citizens in dialogues about AI's applications and risks. Such involvement could significantly influence democratic decision-making processes.

Our study also showed that age negatively correlates with the perceived need for auditing. This trend could stem from older individuals' prior experiences, which may have led to a certain disillusionment with auditing practices in various societal domains. Additionally, our data indicates that male participants are more inclined to support the registration of RBI systems in a public database. Future research could explore the underlying reasons for these demographic differences by expanding on these findings. For instance, it would be insightful to examine how generational experiences and gender perspectives shape attitudes towards transparency and oversight in the realm of AI technologies. This exploration could provide valuable insights into tailoring communication and policy strategies to address different demographic groups' diverse concerns and viewpoints regarding AI governance.

\hypertarget{the-role-of-discrimination-perceptions}{%
\subsection{The Role of Discrimination
Perceptions}\label{the-role-of-discrimination-perceptions}}

As outlined above, discrimination perceptions regarding RBI technology
only mediate some effects. However, looking at the
direct effects of citizens' discrimination perceptions on the demand for
different types of regulation is also worthwhile.

While perceptions of RBI's discriminatory impacts are not directly
affected by the contextual configurations of RBI systems, they are
associated with support for regulatory measures. A general
awareness of discrimination related to RBI systems impacts regulatory
demands. This effect is more reflective of an overarching attitude
towards RBI's discriminatory implications rather than its specific
applications. This aligns with Kieslich et al.'s \cite{Kieslich.2023} findings,
who highlight that ethical issue awareness intensifies political
engagement with AI.

Notable disparities emerge when scrutinizing the impact of the three dependent variables. The most pronounced effects are observed in the context of supporting a complete ban on RBI systems. In contrast, while
the impact of advocating for audits and registrations is significant, it
is comparatively modest. Therefore, the perception of RBI as
discriminatory propels citizens towards demanding more stringent
regulations, specifically advocating for outright bans. Conversely,
milder forms of regulation, such as mandatory database registration and
auditing, although positively correlated with higher discrimination
perceptions, elicit weaker reactions. As discrimination represents a serious concern in the context of human rights, it is plausible that people tend
to opt for an outright ban rather than a mild form of regulation if
they think that problematic discrimination is likely to happen when deploying RBI.
This is also in line with the notion of unacceptable risk as classified
by the EU AI Act. In this argumentation, some technologies are too
risky to be implemented into society. It seems like if people perceive
this risk, then they also opt for setting limits to the introduction of
AI technologies.

This is a promising finding for the AI ethics community as it shows that
awareness of discriminatory risks regarding RBI can, in fact, impact the
support of regulatory approaches. However, in light of the findings
above, this attitude is not nuanced in the sense that it is tied to the
context of the use of RBI but rather reflects a general attitude towards RBI.
Further, as elaborated above, trust in AI or in law enforcement can
weaken discrimination perceptions. Thus, it remains an open question
how discrimination perceptions can become more nuanced and reflective, leading to a deliberate and careful demand for regulation. Future research should try to fill this
gap and illuminate the impact of additional predictors of regulatory demands. One potential
factor frequently discussed in this regard could be
strengthening AI literacy, especially regarding the social impacts of
the technology \cite{Long.2020c, OeldorfHirsch.2023}.

\hypertarget{conclusion}{%
\section{Conclusion}\label{conclusion}}

In this study, we tapped into German citizens' perceptions of the
regulation of RBI systems. RBI systems are classified in the draft of
the EU AI Act as an unacceptable risk or high risk depending on the
temporal aspect and purpose of its application. However, as these
classifications are drafted top-down by an expert commission, we
explored the opinions of the populace and researched factors that lead
to support regulatory measures regarding the use of RBI in
different contexts -- thus, bringing society in the loop \cite{Rahwan.2018}. We
were especially interested in aspects of trust in law enforcement as
well as in AI technology and discrimination perceptions of RBI systems.
Our results from a factorial survey study showed that when it comes 
to regulation and discrimination perception, citizens do not distinguish 
between real-time and post hoc use and different purposes of RBI use. 
The fine-grained distinction made by the EU AI Act is not reflected in 
citizens' opinion -- which questions the ability of citizens to engage in 
these detailed discussions about the deployment conditions of specific 
AI applications.

However, we found that general discrimination perceptions of the use of
RBI impacts the demand for stronger regulation. If citizens show a high
awareness of the discriminatory impact of RBI, they want these technologies 
to be banned -- in terms of the EU AI Act classified as
an unacceptable risk. This holds true, even if the technology is --
according to the latest version of the AI Act -- ``only'' deemed as high
risk. Other counter-measures like stronger auditing or registration of
RBI in a public database also finds support but are not as strongly
positively influenced as the other dimensions. Interestingly, we also
found that trust plays a significant role in smoothing regulatory
demands. If citizens trust in law enforcement or in AI, they show less
tendency to opt for a ban on RBI systems.

This study has shown the potential for public demands in terms of that
awareness in the citizenry can lead to a demand for stronger
regulation. At the same time,  it also underscores the limitations of public opinion in regard to its inability to detect fine-grained distinctions between different use cases of a risky technology. Despite these limitations, the importance of incorporating citizens' perspectives in AI governance is emphasized, especially when AI is implemented in public domains. This inclusion ensures that the voices of the citizenry are considered in shaping policies for technologies that significantly impact society.

%\hypertarget{references}{%
%\section*{References}\label{references}}

\hypertarget{ethics-statement}{%
\section*{Ethics Statement}\label{ethics-statement}}

In conducting this research, we adhered to the highest standards of
ethical integrity and responsibility. We ensured that all participants
were fully informed about the nature and purpose of the research and
provided their informed consent. Participant confidentiality and data
privacy were rigorously maintained throughout the study. Ethical
guidelines, including those pertaining to non-discrimination, fairness,
and respect for individuals, were strictly followed. The research
methods were designed to minimize potential harm or discomfort to
participants. Any conflicts of interest were disclosed and managed
appropriately. The questionnaire was reviewed by two academic scholar invited by the SoSci panel; both reviewers did not raise ethical concerns in regard to our study. 

\hypertarget{conflict-interest}{%
\section*{Conflicts of Interest}\label{conflict-interest}}

The authors declare that they have no known competing financial
interests or personal relationships that could have appeared to
influence the work reported in this paper.

\hypertarget{declaration-genAI}{%
\section*{Declaration of Generative AI in Scientific Writing}\label{declaration-genAI}}

During the preparation of this work, the authors used ChatGPT in order
to produce and adjust R and Markdown code for the statistical analysis
and the reproducible manuscript. After using this
tool/service, the authors reviewed and edited the content as needed and
take full responsibility for the content of the publication.

\bibliographystyle{ACM-Reference-Format}
\bibliography{references}

\newpage

\hypertarget{appendix}{%
\section*{Appendix}\label{appendix}}
\addcontentsline{toc}{section}{Appendix}

\hypertarget{vignette-original-german-wording}{%
\subsection*{Vignette original (German)
wording}\label{vignette-original-german-wording}}
\addcontentsline{toc}{subsection}{Vignette original (German) wording}

\hypertarget{real-time}{%
\subsubsection*{Real-Time}\label{real-time}}
\addcontentsline{toc}{subsubsection}{Real-Time}

\textbf{Der Einsatz von biometrischer Fernidentifikation zur 1)
Echtzeit-Fahndung nach Schwerverbrecherinnen und Schwerverbrechern 2) Sicherung von
Großveranstaltungen}

Ein derzeit von der Politik diskutiertes Thema ist der Einsatz von
KI-basierten Systemen zur biometrischen Identifizierung von Personen,
sogenannte Systeme zur \textbf{biometrischen Fernidentifikation}. Damit
sind Computeranwendungen gemeint, die physische Merkmale von Personen
(z. B. Gesichter) analysieren und damit konkreten Personen zuordnen
können.

Biometrische Fernidentifikation kann von Strafverfolgungsbehörden
\textbf{in Echtzeit} eingesetzt werden, um Personen zu identifizieren
und zu verfolgen, die möglicherweise in kriminelle Aktivitäten
verwickelt sind.

Sie funktioniert, indem Kameras oder andere Sensoren Bilder von
Gesichtern, Fingerabdrücken oder anderen biometrischen Merkmalen einer
Person aufnehmen und diese Bilder mit einer Datenbank bekannter Personen
vergleichen.

Die Technologie verwendet Algorithmen und Künstliche Intelligenz, um die
aufgenommenen Bilder zu analysieren und Übereinstimmungen mit der
Datenbank zu finden. Dies ermöglicht es der Strafverfolgung, potenzielle
Verdächtige schnell zu identifizieren und ihre Bewegungen zu verfolgen,
selbst wenn sie sich in einem öffentlichen Bereich aufhalten oder
versuchen, sich der Erkennung zu entziehen.

Ein relevantes Anwendungsgebiet von biometrischer Fernidentifikation ist
der Einsatz zur 1) \textbf{Fahndung nach Schwerverbrecherinnen und Schwerverbrechern 2)
Sicherung von Großerveranstaltungen (z. B. Sportveranstaltungen)}.

\hypertarget{post-hoc-analysis}{%
\subsubsection*{Post hoc analysis}\label{post-hoc-analysis}}
\addcontentsline{toc}{subsubsection}{Post hoc analysis}

\textbf{Der Einsatz von biometrischer Fernidentifikation 1) zur
Erkennung von Schwerverbrecherinnen und Schwerverbrechern} \textbf{2) nach Zwischenfällen bei
Großveranstaltungen}

Ein derzeit von der Politik diskutiertes Thema ist der Einsatz von
KI-basierten Systemen zur biometrischen Identifizierung von Personen,
sogenannte Systeme zur \textbf{biometrischen Fernidentifikation}. Damit
sind Computeranwendungen gemeint, die physische Merkmale von Personen
(z. B. Gesichter) analysieren und damit konkreten Personen zuordnen
können.

Biometrische Fernidentifikation kann von Strafverfolgungsbehörden
\textbf{im Nachgang von kriminellen Ereignissen} eingesetzt werden, um
Personen zu identifizieren und zu verfolgen, die möglicherweise in diese
verwickelt waren.

Sie funktioniert, indem Kameras oder andere Sensoren Bilder von
Gesichtern, Fingerabdrücken oder anderen biometrischen Merkmalen einer
Person aufnehmen und diese Bilder mit einer Datenbank bekannter Personen
vergleichen.

Die Technologie verwendet Algorithmen und Künstliche Intelligenz, um die
aufgenommenen Bilder zu analysieren und Übereinstimmungen mit der
Datenbank zu finden. Dies ermöglicht es der Strafverfolgung, potenzielle
Verdächtige schnell zu identifizieren und ihre Bewegungen zu verfolgen,
selbst wenn sie sich in einem öffentlichen Bereich aufhielten oder
versuchten, sich der Erkennung zu entziehen.

Ein relevantes Anwendungsgebiet von biometrischer Fernidentifikation ist
der Einsatz zur \textbf{1) Erkennung von Schwerverbrecherinnen und Schwerverbrechern 2)
Sicherung von Großerveranstaltungen (z. B. Sportveranstaltungen).}

\newpage

\hypertarget{vignette-translated-wording}{%
\subsection*{Vignette translated
wording}\label{vignette-translated-wording}}
\addcontentsline{toc}{subsection}{Vignette translated wording}

\hypertarget{real-time-1}{%
\subsubsection*{Real-Time}\label{real-time-1}}
\addcontentsline{toc}{subsubsection}{Real-Time}

\textbf{The use of remote biometric identification for 1) real-time
searches for serious criminals 2) securing public events}.

A topic currently being discussed by politicians is the use of AI-based
systems for the biometric identification of persons, so-called systems
for \textbf{biometric remote identification}. This refers to computer
applications that can analyse physical characteristics of persons
(e.g.~faces) and thus assign them to concrete persons.

Remote biometric identification can be used by law enforcement agencies
\textbf{in real time} to identify and track individuals who may be
involved in criminal activity.

It works by cameras or other sensors capturing images of a person's
face, fingerprints or other biometric characteristics and comparing
these images to a database of known individuals.

The technology uses algorithms and artificial intelligence to analyse
the captured images and find matches with the database. This allows law
enforcement to quickly identify potential suspects and track their
movements, even if they are in a public area or trying to evade
detection.

A relevant application of remote biometric identification is its use for
1) \textbf{searching for serious criminals 2) securing public
events (e.g.~sporting events)}.

\hypertarget{post-hoc-analysis-1}{%
\subsubsection*{Post hoc analysis}\label{post-hoc-analysis-1}}
\addcontentsline{toc}{subsubsection}{Post hoc analysis}

\textbf{The use of remote biometric identification 1) for the detection
of serious criminals} \textbf{2) after incidents at public events}.

A topic currently being discussed by politicians is the use of AI-based
systems for the biometric identification of persons, so-called systems
for \textbf{biometric remote identification}. This refers to computer
applications that can analyse physical characteristics of persons
(e.g.~faces) and thus assign them to concrete persons.

Remote biometric identification can be used by law enforcement
\textbf{in the aftermath of criminal events} to identify and track
individuals who may have been involved.

It works by cameras or other sensors taking images of a person's face,
fingerprints or other biometric characteristics and comparing these
images with a database of known individuals.

The technology uses algorithms and artificial intelligence to analyse
the captured images and find matches with the database. This allows law
enforcement to quickly identify potential suspects and track their
movements, even if they were in a public area or trying to evade
detection.

A relevant application of remote biometric identification is its use for
\textbf{1) recognition of serious criminals 2) securing public
events (e.g.~sporting events).}
\setlength{\parindent}{0in}

\end{document}